\documentclass[twoside,leqno,twocolumn]{article}

\newif\ifFull
\Fulltrue

\usepackage[letterpaper]{geometry}

\usepackage{ltexpprt}
\usepackage{t1enc}
\usepackage[utf8]{inputenc}
\usepackage{numprint}
\npdecimalsign{.} 
\usepackage{hyperref}
\usepackage{xspace}
\usepackage{doi}
\usepackage{enumerate}
\usepackage{booktabs}

\usepackage{amsmath}
\usepackage{amssymb}

\newcommand{\etal}{et~al.\xspace}

\newcommand{\mytitle}{WeGotYouCovered: The Winning Solver from the PACE 2019 Implementation Challenge, Vertex Cover Track}
\usepackage{changes}
\newcommand{\AlgName}[1]{\textsf{#1}}

\begin{document}
\title{\mytitle\thanks{
    The research leading to these results has received funding from the European Research Council under the European Community's Seventh Framework Programme (FP7/2007-2013) /ERC grant agreement No. 340506}}
\author{Demian Hespe\thanks{Karlsruhe Institute of Technology, Karlsruhe, Germany} \and Sebastian Lamm\thanks{Karlsruhe Institute of Technology, Karlsruhe, Germany} \and Christian Schulz\thanks{University of Vienna, Faculty of Computer Science, Austria} \and Darren Strash\thanks{Hamilton College, New York, USA}}

\date{}

\maketitle
\begin{abstract}
We present the winning solver of the PACE 2019 Implementation Challenge, Vertex Cover Track. The minimum vertex cover problem is one of a handful of problems for which \emph{kernelization}---the repeated reducing of the input size via \emph{data reduction rules}---is known to be highly effective in practice. Our algorithm uses a portfolio of techniques, including an aggressive kernelization strategy, local search, branch-and-reduce, and a state-of-the-art branch-and-bound solver. Of particular interest is that several of our techniques were \emph{not} from the literature on the vertex over problem: they were originally published to solve the (complementary) maximum independent set and maximum clique problems. 

Aside from illustrating our solver's performance in the PACE 2019 Implementation Challenge, our experiments provide several key insights not yet seen before in the literature.
First, kernelization can boost the performance of branch-and-bound clique solvers enough to outperform branch-and-reduce solvers. Second, local search can significantly boost the performance of branch-and-reduce solvers. And finally, somewhat surprisingly, kernelization can sometimes make branch-and-bound algorithms perform \emph{worse} than running branch-and-bound alone.
\end{abstract}

\section{Introduction}

A \emph{vertex cover} of a graph $G=(V,E)$ is a set of vertices $S\subseteq V$ of $G$ such that every edge of $G$ has at least one of member of $S$ as an endpoint (i.e., $\forall (u,v) \in E\,\, [u\in S \textrm{ or } v \in S]$).
The minimum vertex cover problem---that of computing a vertex cover of minimum cardinality---is a fundamental NP-hard problem, and has applications spanning many areas. These include computational biology~\cite{cheng2008prediction}, classification~\cite{gottlieb2014efficient}, mesh rendering~\cite{sander2008efficient}, and many more through its complementary problems~\cite{gnntbdmlisaac13,gardiner-docking-2000,harary-clique-1957,zaki-ecommerce-97}.

Complementary to vertex covers are independent sets and cliques. An independent set is a set of vertices $I\subseteq V$, all pairs of which are not adjacent, and an clique is a set of vertices $K\subseteq V$ all pairs of which are adjacent. A maximum independent set (maximum clique) is an independent set (clique) of maximum cardinality. The goal of the maximum independent set problem (maximum clique problem) is to compute a maximum independent set (maximum clique).

Many techniques have been proposed for solving these problems, and papers in the literature usually focus on one of these problems in particular. However, all of these problems are equivalent: a
minimum vertex cover $C$ in $G$ is the complement of a maximum independent set $V\setminus C$ in $G$, which is a maximum clique $V\setminus C$ in $\overline{G}$. Thus, an algorithm that solves one of these problems can be used to~solve~the~others.
To win the PACE 2019 Implementation Challenge, we deployed a portfolio of solvers, using techniques from the literature on all three problems. These include data reduction rules and branch-and-reduce for the minimum vertex cover problem~\cite{akiba-tcs-2016}, iterated local search for the maximum independent set problem~\cite{andrade-2012}, and a state-of-the-art branch-and-bound maximum clique solver~\cite{DBLP:journals/cor/LiJM17}.

\paragraph*{Our Results.}
In this paper, we describe our techniques and solver in detail and analyze the results of our experiments on the data sets provided by the challenge.
Not only do our experiments illustrate the power of the techniques spanning the literature, they also provide several new insights not yet seen before.
In particular, kernelization followed by branch-and-bound can outperform branch-and-reduce solvers; seeding branch-and-reduce by an initial solution from local search can significantly boost its performance; and, somewhat surprisingly, kernelization is sometimes counterproductive: branch-and-bound algorithms can perform significantly worse on the kernel than on the original input graph.

\paragraph*{Organization.}
We first briefly describe related work in Section~\ref{sec:related_work}. Then in Section~\ref{sec:techniques} we outline each of the techniques that we use, and in Section~\ref{sec:puttingtogether} finally describe how we combine all of the techniques in our final solver that scored the most points in the PACE 2019 Implementation Challenge. Lastly, in Section~\ref{sec:experiments} we perform an experimental evaluation to show the impact of the components used on the final number of instances solved during the challenge.

\section{Preliminaries}
\label{sec:preliminaries}
We work with an undirected graph $G = (V,E)$ where $V$ is a set of $n$ vertices and $E\subset \{\{u,v\}\mid u,v\in V\}$ is a set of $m$ edges. The open neighborhood of a vertex $v$, denoted $N(v)$, is the set of all vertices $w$ such that $(v,w)\in E$. We further denote the closed neighborhood by $N[v]=N(v)\cup\{v\}$. We similarly define the open and closed neighborhoods of a set of vertices $U$ to be $N(U) = \bigcup_{u\in U}N(u)$ and $N[U] = N(U) \cup U$, respectively. The set of vertices of distance $d$ of a vertex $u$ is denoted by $N^d(u)$, where $N^2(u)$ is called the \emph{two-neighborhood} of $u$. Lastly, for vertices $S\subseteq V$, the induced subgraph $G[S]\subseteq G$ is the graph on the vertices in $S$ with edges in $E$ between vertices in $S$.

\section{Related Work}
\label{sec:related_work}
Research results in the area can be found through work on the minimum vertex cover problem and its complementary maximum clique and independent set problems, and can often be categorized depending on the angle of attack. For exact exponential (theoretical) algorithms, the maximum independent set problem is canonically studied, for parameterized algorithms, the minimum vertex cover problem is studied, and the maximum clique problem is normally solved exactly in practice (though there are recent exceptions). However, these problems are only \emph{trivially} different --- techniques for solving one problem require only subtle modifications to solve the other two.

\paragraph*{Exponential-time Algorithms.}
The maximum independent set problem is most often considered when designing exact (exponential-time) algorithms, and much research has be devoted to reducing the base of the exponential running time. A primary technique is  to develop rules to modify the graph, removing or contracting subgraphs that can be solved simply, which reduces the graph to a smaller instance. These rules are referred to as \emph{data reduction rules} (often simplified to \emph{reduction rules} or \emph{reductions}).
Reduction rules have been used to reduce the running time of the brute force $O(n^22^n)$ algorithm to the $O(2^{n/3})$ time algorithm of Tarjan and Trojanowski~\cite{tarjan-1977}, and to the current best polynomial space algorithm with running time of $O^*(1.1996^n)$ by Xiao and Nagamochi~\cite{xiao2017exact}. 

The reduction rules used for these algorithms are often staggeringly simple, including \emph{pendant vertex removal}, \emph{vertex folding}~\cite{chen1999} and \emph{twin} reductions~\cite{Xiao201392}, which eliminate nearly all vertices of degree three or less from the graph. 
These algorithms apply reductions during recursion, only branching when the graph can no longer be reduced~\cite{fomin-2010}, and are referred to as \emph{branch-and-reduce} algorithms. Further techniques used to accelerate these algorithms include \emph{branching rules}~\cite{kneis2009fine,fomin2009measure} which eliminate unnecessary branches from the search tree, as well as faster exponential-time algorithms for graphs of small maximum degree~\cite{xiao2017exact}.

\paragraph*{Parameterized Algorithms.}
For parameterized algorithms, we now turn to the minimum vertex cover problem. The most efficient algorithms for computing a minimum vertex cover in both theory and practice repeatedly apply data reduction rules to obtain a (hopefully) much smaller problem instance. If this smaller instance has size bounded by a function of some parameter, it's~called~a~\emph{kernel}, and producing a polynomially-sized kernel gives a fixed-parameter tractable in the chosen parameter. Reductions are surprisingly effective for the minimum vertex cover problem. In particular, letting $k$ be the size of a minimum vertex cover, the well-known crown reduction rule produces a kernel of size $3k$~\cite{chor2005linear} and the LP-relaxation reduction due to Nemhauser and Trotter~\cite{nemhauser-1975}, produces a kernel of size $2k$~\cite{chen1999}. Chen et al.~\cite{chen2010improved} developed the current best parameterized algorithm for minimum vertex cover, giving a branch-and-reduce algorithm with running time $O(1.2738^k +kn)$ and polynomial space.
For more information on the history of vertex cover kernelization, see the recent survey by Fellows et al.~\cite{fellows2018known}.

\paragraph*{Exact Algorithms in Practice.}
The most efficient maximum clique solvers use a branch-and-bound search with advanced vertex reordering strategies and pruning (typically using approximation algorithms for graph coloring, MaxSAT~\cite{li-maxsat-2013} or constraint satisfaction). The long-standing canonical algorithms for finding the maximum clique are the MCS algorithm by Tomita et al.~\cite{tomita-recoloring} and the bit-parallel algorithms of San Segundo et al.~\cite{segundo-recoloring,segundo-bitboard-2011}. However, recently Li et al.~\cite{DBLP:journals/cor/LiJM17} introduced the MoMC algorithm, which uses incremental MaxSAT logic to achieve speed ups of up to \numprint{1000} over MCS. Experiments by Batsyn et al.~\cite{batsyn-mcs-ils-2014} show that MCS can be sped up significantly by giving an initial solution found through local search. However, even with these state-of-the-art algorithms, graphs on thousands of vertices remain intractable. For example, a difficult graph on \numprint{4000} required 39 wall-clock hours in a highly-parallel MapReduce cluster, and is estimated to require over a year of sequential computation~\cite{xiang-2013}. Recent clique solvers for sparse graphs investigate applying simple data reduction rules, using an initial clique given by some inexact method~\cite{verma2015solving,sansegundo2016a,chang2019efficient}. However, these techniques rarely work on dense graphs, such as the complement graphs that we consider here.
A thorough discussion of many results in clique finding can be found in the survey of Wu and Hao~\cite{wu-hao-2015}.

Data reductions have been successfully applied in practice to solve many problems that are intractable with general algorithms. Butenko et al.~\cite{butenko-2002,butenko-correcting-codes-2009} were the first to show that simple reductions could be used to compute exact maximum independent sets on graphs with hundreds vertices for graphs derived from error-correcting codes. Their algorithm works by first applying \emph{isolated clique removal} reductions, then solving the remaining graph with a branch-and-bound algorithm. Later, Butenko and Trukhanov~\cite{butenko-trukhanov} introduced the \emph{critical independent set} reduction, which was able to solve graphs produced by the Sanchis graph generator.
Larson~\cite{larson-2007} later proposed an algorithm to find a \emph{maximum} critical independent set, but in experiments it proved to be slow in practice~\cite{strash2016power}.
Iwata~\etal~\cite{iwata-2014} then showed how to remove a large collection of vertices from a maximum matching all at once; however, it is not known if these reductions are equivalent.

For the minimum vertex cover problem, it has long been known that two such simple reductions, called \emph{pendant vertex removal} and \emph{vertex folding}, are particularly effective in practice. However, two seminal experimental works explored the efficacy of further reductions. Abu-Khzam et al.~\cite{abu-khzam-2007} showed that \emph{crown reductions} are as effective (and sometimes faster) in practice than performing the LP relaxation reduction (which, as they show in the paper, removes crowns) on graphs. We briefly note that critical independent sets, together with their neighborhoods, are in fact crowns, and thus in some ways the work of Butenko and Trukhanov~\cite{butenko-trukhanov} replicates that by Abu-Khzam et al.~\cite{abu-khzam-2007}, though their experiments are run on different graphs.

Later, Akiba and Iwata~\cite{akiba-tcs-2016} showed that an extensive collection of advanced data reduction rules (together with branching rules and lower bounds for pruning search) are also highly effective in practice. Their algorithm finds exact minimum vertex covers on a corpus of large social networks with hundreds of thousands of vertices or more in mere seconds. More details on the reduction rules follow in Section~\ref{sec:techniques}.

We briefly note that we considered other reduction techniques that emphasize fast computation at the cost of a larger (irreducible) graph~\cite{chang2017computing,strash2016power,DBLP:conf/alenex/Hespe0S18}; however, we did not find them as effective as Akiba and Iwata~\cite{akiba-tcs-2016} for exactly solving difficult instances. This is somewhat expected, however, since these techniques are optimized to produce fast high-quality solutions when combined with inexact methods such as local search.

\section{Techniques}
\label{sec:techniques}
We now describe techniques that we use in~our~solver.
\subsection{Kernelization.}
The most efficient algorithms for computing a minimum vertex cover in both theory and practice use \emph{data reduction rules} to obtain a much smaller problem instance. If this smaller instance has size bounded by a function of some parameter, it's~called~a~\emph{kernel}. 

We use an extensive (though not exhaustive) collection of data reduction rules whose efficacy was studied by Akiba and Iwata~\cite{akiba-tcs-2016}. To compute a kernel, Akiba and Iwata~\cite{akiba-tcs-2016} apply their
reductions~$r_1, \dots, r_j$ by iterating over all reductions and trying to
apply the current reduction $r_i$ to all vertices. If $r_i$ reduces at
least one vertex, they restart with reduction~$r_1$. When reduction~$r_j$ 
is executed, but does not reduce any vertex, all reductions have been applied
exhaustively, and a kernel is found. Following their study we order the reductions
as follows: degree-one vertex (i.e., pendant) removal, unconfined vertex removal~\cite{Xiao201392}, a well-known linear-programming 
relaxation~\cite{iwata-2014,nemhauser-1975} (which, consequently, removes crowns~\cite{abu-khzam-2007}),  vertex folding~\cite{chen1999}, and twin, funnel, and desk reductions~\cite{Xiao201392}.

To be self-contained, we now give a brief description of those reductions, in order of increasing complexity. Each reduction allows us to choose vertices that are either in some minimum vertex cover, or for which we can locally choose a vertex in a minimum vertex cover after solving the remaining graph, by following simple rules. If a minimum vertex cover  is found in the kernel, then each reduction may be undone, producing a minimum vertex cover in the original graph. Refer to Akiba and Iwata~\cite{akiba-tcs-2016} for a more thorough discussion, including implementation details. Our implementation of the reductions is an adaptation of Akiba and Iwata's original code. \\

\noindent\textbf{Pendant vertices:} Any vertex $v$ of degree one, called a \emph{pendant}, then its neighbor is in some minimum vertex cover, therefore $v$ and its neighbor $u$ can be removed from $G$. \\

\noindent\textbf{Vertex folding:} For a vertex $v$ with degree 2 whose neighbors $u$ and $w$ are not adjacent, either $v$ is in some minimum vertex cover, or both $u$ and $w$ are in some minimum vertex cover. Therefore, we can contract $u$, $v$, and $w$ to a single vertex $v'$ and decide which vertices are in the vertex cover after computing a minimum vertex cover on the reduced graph. \\

\noindent\textbf{Linear Programming Relaxation:}
First introduced by Nemhauser and Trotter~\cite{nemhauser-1975} for the vertex packing problem, they present a linear programming relaxation with a half-integral solution (i.e., using only values 0, 1/2, and 1) which can be solved using bipartite matching. Their relaxation may be formulated for the minimum vertex cover problem as follows: minimize $\sum_{v\in V}{x_v}$, such at for each edge $(u, v) \in E$, $x_u + x_v \geq 1$ and for each vertex $v \in V$, $x_v \geq 0$. There is a minimum vertex cover containing no vertices with value $1$, and therefore their neighbors are added to the solution and removed together with the vertices from the graph. We use the further improvement from Iwata et al.~\cite{iwata-2014}, which computes a solution whose half-integral part is minimal. \\

\noindent\textbf{Unconfined~\cite{Xiao201392}:} Though there are several definitions of an \emph{unconfined} vertex in the literature, we use the simple one from Akiba and Iwata~\cite{akiba-tcs-2016}. A vertex $v$ is \emph{unconfined} when determined by the following simple algorithm. First, initialize $S = \{v\}$. Then find a $u \in N(S)$ such that $|N(u) \cap S| = 1$ and $|N(u) \setminus N[S]|$ is minimized. If there is no such vertex, then $v$ is confined. If $N(u) \setminus N[S] = \emptyset$, then $v$ is unconfined.  If $N(u)\setminus N[S]$ is a single vertex $w$, then add $w$ to $S$ and repeat the algorithm. Otherwise, $v$ is confined. Unconfined vertices can be removed from the graph, since there always exists a minimum vertex cover that contains unconfined vertices. \\

\noindent\textbf{Twin~\cite{Xiao201392}:} Let $u$ and $v$ be vertices of degree 3 with $N(u) = N(v)$. If $G[N(u)]$ has edges, then add $N(u)$ to the minimum vertex cover and remove $u$, $v$, $N(u)$, $N(v)$ from $G$. Otherwise, some $u$ and $v$ may belong to some minimum vertex cover. We still remove $u$, $v$, $N(u)$ and $N(v)$ from $G$, and add a new gadget vertex $w$ to $G$ with edges to $u$'s two-neighborhood (vertices at a distance 2 from $u$). If $w$ is in the computed minimum vertex cover, then $u$'s (and $v$'s) neighbors are in some minimum vertex cover, otherwise $u$ and $v$ are in a minimum vertex cover.\\

\noindent\textbf{Alternative:} Two sets of vertices $A$ and $B$ are set to be \emph{alternatives} if $|A| = |B| \geq 1$ and there exists an minimum vertex cover $C$ such that $C\cap(A\cup B)$ is either $A$ or $B$. Then we remove $A$ and $B$ and $C = N(A)\cap N(B)$ from $G$ and add edges from each $a \in N(A)\setminus C$ to each $b\in N(B)\setminus C$.
Then we add either $A$ or $B$ to $C$, depending on which neighborhood has vertices in $C$. Two structures are detected as alternatives. First, if $N(v)\setminus \{u\}$ induces a complete graph, then $\{u\}$ and $\{v\}$ are alternatives (a \emph{funnel}). Next, if there is a cordless 4-cycle $a_1b_1a_2b_2$ where each vertex has at least degree 3. Then sets $A=\{a_1, a_2\}$ and $B=\{b_1, b_2\}$ are alternatives (called a \emph{desk}) when $|N(A) \setminus B| \leq 2$, $|N(A) \setminus B| \leq 2$, and $N(A) \cap N(B) = \emptyset$. 

\subsection{Branch-and-Reduce.}
Branch-and-reduce is a paradigm that intermixes data reduction rules and branching. We use the algorithm of Akiba and Iwata, which exhaustively applies their full suite of reduction rules before branching, and includes a number of advanced branching rules as well as lower bounds to prune search. 

\paragraph*{Branching.}
When branching, a vertex of maximum degree is chosen for inclusion into the vertex cover. Mirrors and satellites are detected when branching, in order to eliminate branching on certain vertices. A \emph{mirror} of a vertex $v$ is a vertex $u\in N^2(v)$ such that $N(v)\setminus N(u)$ is a clique or empty. Fomin et al.~\cite{fomin2009measure} show that either the mirrors of $v$ or $N(v)$ is in a minimum vertex cover, and we can therefore branch on all mirrors at once. This branching prevents branching on mirrors individually and decreases the size of the remaining graph (and thus the depth of the search tree). A \emph{satellite} of a vertex $v$ is a vertex $u\in N^2(v)$ such that there exists a vertex $w\in N(v)$ such that $N(w)\setminus N[v] = \{u\}$. If a vertex has no mirrors, then either $v$ is in a minimum vertex cover or the neighbors of $v$'s satellites in a minimum vertex cover. Akiba and Iwata~\cite{akiba-tcs-2016} further introduce \emph{packing} branching, maintaining linear inequalities for each vertex included or excluded from the current vertex cover (called \emph{packing constraints}) throughout recursion; when a constraint is violated, further branching can be eliminated.

\paragraph*{Lower Bounds.} We briefly remark that Akiba and Iwata~\cite{akiba-tcs-2016} implement lower bounds to prune the search space. Their lower bounds are based on clique cover, the LP relaxation, and cycle covers (see their paper for further details). The final lower bound used for pruning is the maximum of these three.

\subsection{Branch-and-Bound.} Experiments by Strash~\cite{strash2016power} show that the full power of branch-and-reduce is only needed \emph{very rarely} in real-world instances; kernelization followed by a standard branch-and-bound solver is sufficient for many real-world instances. Furthermore, branch-and-reduce does not work well on many synthetic benchmark instances, where data reduction rules are ineffective~\cite{akiba-tcs-2016}, and instead add significant overhead to branch-and-bound. We use a state-of-the-art branch-and-bound maximum clique solver (MoMC) by Li et al.~\cite{DBLP:journals/cor/LiJM17}, which uses incremental MaxSAT reasoning to prune search, and a combination of static and dynamic vertex ordering to select the vertex for branching. We run the clique solver on the complement graph, giving a maximum independent set from which we derive a minimum vertex cover. In preliminary experiments, we found that a kernel can sometimes be harder for the solver than the original input; therefore, we run the algorithm on both the kernel and on the original graph.

\subsection{Iterated Local Search.}
Batsyn et al.~\cite{batsyn-mcs-ils-2014} showed that if branch-and-bound search is primed with a high-quality solution from local search, then instances can be solved up to thousands of times faster. 
We use the iterated local search algorithm by Andrade et al.~\cite{andrade-2012} to prime the \emph{branch-and-reduce} solver with a high-quality initial solution. To the best of our knowledge, this has not been tried before. Iterated local search was originally implemented for the maximum independent set problem, and is based on the notion of $(j,k)$-swaps. A $(j,k)$-swap removes $j$ nodes from the current solution and inserts $k$ nodes. The authors present a fast linear-time implementation that, given a maximal independent set, can find a $(1,2)$-swap or prove that none exists. Their algorithm applies $(1,2)$-swaps until reaching a local maximum, then perturbs the solution and repeats. We implemented the algorithm to find a high-quality solution on \emph{the kernel}. Calling local search on the kernel has been shown to produce a high-quality solution much faster than without kernelization~\cite{chang2017computing,dahlum2016accelerating}.

\section{Putting it all Together}
\label{sec:puttingtogether}
Our algorithm first runs a preprocessing phase, followed by 4 phases of solvers.

\begin{description}
\item[Phase 1. (Preprocessing)] Our algorithm starts by computing a kernel of the graph using the reductions by Akiba and Iwata~\cite{akiba-tcs-2016}. 
From there we use iterated local search to produce a high-quality solution $S_{\textrm{init}}$ on the (hopefully smaller) kernel. 
\item[Phase 2. (Branch-and-Reduce, short)]
We prime a branch-and-reduce solver with the initial solution $S_{\textrm{init}}$ and run it with a short time limit.
\item[Phase 3. (Branch-and-Bound, short)]
If Phase 2 is unsuccessful, we run the MoMC~\cite{DBLP:journals/cor/LiJM17} clique solver on the complement of the kernel, also using a short time limit\footnote{Note that repeatedly checking the time can slow down a highly optimized branch-and-bound solver considerably; we therefore simulate time checking by using a limit on the number of branches.}. Sometimes kernelization can make the problem harder for MoMC. Therefore, if the first call was unsuccessful we also run MoMC on the complement of the original (unkernelized) input with the same short time limit.

\item[Phase 4. (Branch-and-Reduce, long)]
If we have still not found a solution, we run branch-and-reduce on the kernel using initial solution $S_{\textrm{init}}$ and a longer time limit. We opt for this second phase because, while most graphs amenable to reductions are solved very quickly with branch-and-reduce (less than a second),
experiments by Akiba and Iwata~\cite{akiba-tcs-2016} showed that other slower instances either finish in at most a few minutes, or take significantly longer---more than the time limit allotted for the challenge. This second phase of branch-and-reduce is meant to catch any instances that still benefit from reductions.

\item[Phase 5. (Branch-and-Bound, remaining time)]
If all previous phases were unsuccessful, we run MoMC on the original (unkernelized) input graph until the end of the time given to the program by the challenge. This is meant to capture only the hardest-to-compute instances.
\end{description}

The algorithm time limits (discussed in the next section) and ordering were carefully chosen so that the overall algorithm outputs solutions of the ``easy'' instances \emph{quickly}, while still being able to solve hard instances.
\section{Experimental Results}
\label{sec:experiments}
We now look at the impact of the algorithmic components on the number of instances solved.
Here, we focus on the public instances  of the PACE 2019 Implementation Challenge, Vertex Cover Track A, obtained from \url{https://pacechallenge.org/files/pace2019-vc-exact-public-v2.tar.bz2}. This set contains 100 instances overall. We also summarize the results comparing against the second and third best competing algorithms on the private instances during the challenge (which can be found at \url{https://pacechallenge.org/2019/} and \url{https://www.optil.io/optilion/problem/3155}). Note that further comparisons are not yet possible, as the private instances have not yet been released. 

\subsection{Methodology and Setup.}
All of our experiments were run on a machine with  four sixteen-core Intel Xeon Haswell-EX E7-8867 processors running at $2.5$ GHz, $1$ TB of main memory, and \numprint{32768} KB of L2-Cache.
The machine runs Debian GNU/Linux 9 and Linux kernel version 4.9.0-9.
All algorithms were implemented in C++11 and compiled with gcc~version 6.3.0 with optimization flag \texttt{-O3}. Our source code is publicly available under the MIT license at~\cite{wegotyoucovered2019}.
Each algorithm was run sequentially with a time limit of 30 minutes---the time allotted to solve a single data set in the PACE 2019 Implementation Challenge. Our primary focus is on the total number of instances solved.
\subsection{Evaluation.}
We now explain the main configuration that we use in our experimental setup.
In the following, \AlgName{MoMC} runs the MoMC clique solver by Li et
al.~\cite{DBLP:journals/cor/LiJM17} on the complement of the input graph;
\AlgName{RMoMC} applies reductions to the input graph exhaustively, and then
runs MoMC on the complement of the resulting kernel; \AlgName{LSBnR} applies
reductions exhaustively, then runs local search to obtain a high-quality
solution on the kernel which is used as a initial bound in the branch-and-reduce
algorithm that is run on the kernel; \AlgName{BnR} applies reductions and then
runs the branch-and-reduce algorithm on the kernel (no local search is used to
improve an initial bound); \AlgName{FullA} is the full algorithm as described in
the previous section, using a short time limit of one second and a long time limit of thirty seconds.

Tables~\ref{tab:detailedresults1} and \ref{tab:detailedresults2} give an overview of the instances that each of the solver solved, including the kernel size, and the minimum vertex cover size for those instances solved by any of the four algorithms.
Overall, \AlgName{MoMC} can solve 30 out of the 100 instances. 
Applying reductions first enables \AlgName{RMoMC} to solve 68 instances. However, curiously, there are two instances (instances 131 and 157) that \AlgName{MoMC} solves, but that \AlgName{RMoMC} can not solve. 
In these cases, kernelization reduced the number of nodes, but \emph{increased} the number of edges. This is due to the \emph{alternative} reduction, which in some cases can create more edges than initially present. This is why we choose to also run MoMC on the unkernelized input graph in \AlgName{FullA} (in order to solve those instances as well).

\begin{figure}
    \centering
    \includegraphics[width=8cm]{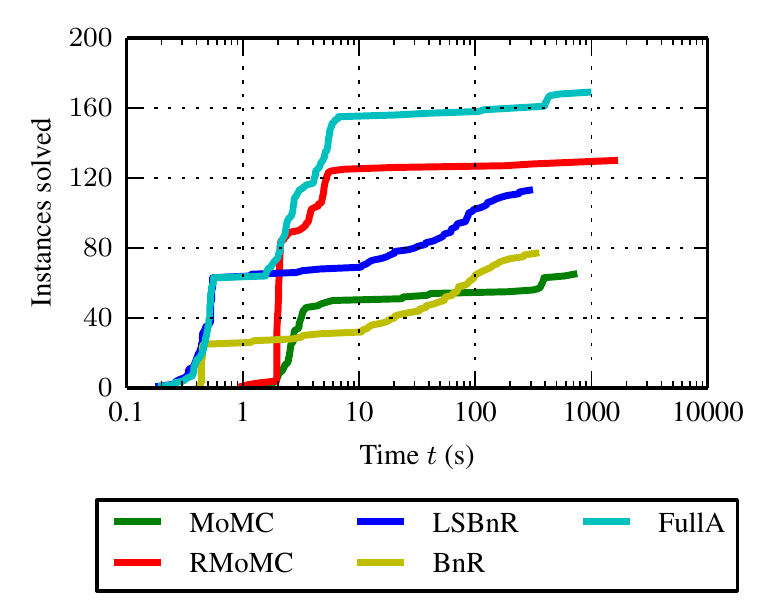}
  \caption{Number of instances solved over time by each algorithm over \emph{all} instances. At each time step $t$, we count each instance solved by the algorithm in at most $t$ seconds.} 
  \vspace*{-.5cm}
  \label{fig:solution_time}
\end{figure}

\AlgName{LSBnR} solves 55 of the 100 instances. Priming the branch-and-reduce algorithm with an initial solution computed by local search has a significant impact: \AlgName{LSBnR} solves 13 more instances than \AlgName{BnR}, which solve 42 instances. In particular, using local search to find an initial bound helps to solve large instances in which the initial kernelization step does not reduce the graph fully. Surprisingly, \AlgName{RMoMC} solves 26 instances that \AlgName{BnR} does not (and even \AlgName{LSBnR} is only able to solve one of these instances). To the best of our knowledge, this is the first time that kernelization followed by branch-and-bound is shown to significantly outperform branch-and-reduce. Our full algorithm \AlgName{FullA} solves 82 of the 100 instances and, as expected, dominates each of the other configurations. This can be further seen from the plot in Figure~\ref{fig:solution_time}, which shows how many instances each algorithm solves over time (this includes all 100 public and 100 private instances of the challenge). Note that \AlgName{LSBnR} and \AlgName{RMoMC} solve more instances in narrow time gaps, due to \AlgName{FullA}'s set up cost and  running multiple algorithms. However, \AlgName{FullA} quickly makes up for this and overtakes all algorithms at approximately eight seconds.
In addition to the 100 public instances, the PACE Implementation Challenge tests all submissions on 100 private instances. Tables~\ref{tab:detailedresultsprivate} and \ref{tab:detailedresults2private} give detailed per instances results on those instances. The results are similar to the results on the private instances. 
On the private instances, \AlgName{MoMC} can solve 35 out of the 100 instances, \AlgName{RMoMC} solves 62, \AlgName{LSBnR} solves 58 and \AlgName{BnR} solves 35 instances.
Our full algorithm \AlgName{FullA} solved 87 of the 100 instances, which is 10 more instances than the second-place submission (\textsf{peaty}~\cite{james_trimble_2019_3082356}, solving 77), and 11 more than the third-place submission (\textsf{bogdan}~\cite{zbogdan_2019_3228802}), solving 76). Our solver dominates these other solvers: with the exception of one graph, our algorithm solves all instances that \textsf{peaty} and \textsf{bogdan} can solve combined. 

We briefly describe these two solvers. The \textsf{peaty} solver uses reductions to compute a problem kernel of the input followed by an unpublished maximum weight clique solver on the complement of each of the connected components of the kernel to assemble a solution. The clique solver is similar to MaxCLQ by Li and Quan~\cite{DBLP:conf/aaai/LiQ10}, but is more general. Local search is used to obtain an initial solution. On the other hand, \textsf{bogdan} implemented a small suite of simple reductions (including vertex folding, isolated clique removal, and degree-one removal) together with a recent maximum clique solver by Szab\'o and Zavalnij~\cite{szabo2018different}. 

Lastly, we note that our choice of using MoMC as our chosen branch-and-bound solver is significant on the private instances. Eight instances solved exclusively by our solver are solved in Phase 5, where MoMC is run until the end of the challenge time limit.
\vfill
\section{Conclusion}
We presented the winning solver of the PACE 2019 Implementation Challenge Vertex Cover Track. Our algorithm uses a portfolio of techniques, including an aggressive kernelization strategy with all known reduction rules, local search, branch-and-reduce, and a state-of-the-art branch-and-bound solver. Of particular interest is that several of our techniques were not from the literature on the vertex over problem: they were originally published to solve the (complementary) maximum independent set and maximum clique problems. Lastly, our experiments show the impact of the different solver techniques on the number of instances solved during the challenge. In particular, the results emphasize that data reductions play an important tool to boost the performance of branch-and-bound, and local search is highly effective to boost the performance of branch-and-reduce.

\subsection*{Acknowledgments.}
We wish to thank the organizers of the PACE 2019 Implementation Challenge for providing us with the opportunity and means to test our algorithmic ideas. We also are indebted to Takuya Akiba and  Yoichi Iwata for sharing their original branch-and-reduce source code\footnote{\url{https://github.com/wata-orz/vertex_cover}}, and to Chu-Min Li, Hua Jiang, and Felip Many\`a for not only sharing---but even open sourcing---their code for MoMC at our request\footnote{\url{https://home.mis.u-picardie.fr/~cli/EnglishPage.html}}. Their solver was of critical importance to our algorithm's success.

\bibliographystyle{plainurl}
\bibliography{references}
\vfill
\begin{table*}
\centering
\caption{Detailed per instance results for public instances. The columns $n$ and $m$ refer to the number of nodes and edges of the input graph, $n'$ and $m'$ refer to the number of nodes and edges of the kernel graph after reductions have been applied exhaustively, and $|VC|$ refers to the size of the minimum vertex cover of the input graph. We list a `\checkmark' when a solver successfully solved the given instance in the time limit, and `-' otherwise.}
\label{tab:detailedresults1}
\begin{tabular}{l@{\hskip 25pt} rrrr|ccccc|rc}
\toprule
inst\# & $n$ &$m$& $n'$& $m'$ & \AlgName{MoMC} & \AlgName{RMoMC} & \AlgName{LSBnR} & \AlgName{BnR} & \AlgName{FullA} & $|VC|$ \\
                \midrule

001 &\numprint{6160}&\numprint{40207}&\numprint{0}&\numprint{0}&-&\checkmark&\checkmark&\checkmark&\checkmark&  \numprint{2586}&\\ 
003 &\numprint{60541}&\numprint{74220}&\numprint{0}&\numprint{0}&-&\checkmark&\checkmark&\checkmark&\checkmark&  \numprint{12190}&\\ 
005 &\numprint{200}&\numprint{819}&\numprint{192}&\numprint{800}&\checkmark&\checkmark&\checkmark&\checkmark&\checkmark&  \numprint{129}&\\ 
007 &\numprint{8794}&\numprint{10130}&\numprint{0}&\numprint{0}&-&\checkmark&\checkmark&\checkmark&\checkmark&  \numprint{4397}&\\ 
009 &\numprint{38452}&\numprint{174645}&\numprint{0}&\numprint{0}&-&\checkmark&\checkmark&\checkmark&\checkmark&  \numprint{21348}&\\ 
011 &\numprint{9877}&\numprint{25973}&\numprint{0}&\numprint{0}&-&\checkmark&\checkmark&\checkmark&\checkmark&  \numprint{4981}&\\ 
013 &\numprint{45307}&\numprint{55440}&\numprint{0}&\numprint{0}&-&\checkmark&\checkmark&\checkmark&\checkmark&  \numprint{8610}&\\ 
015 &\numprint{53610}&\numprint{65952}&\numprint{0}&\numprint{0}&-&\checkmark&\checkmark&\checkmark&\checkmark&  \numprint{10670}&\\ 
017 &\numprint{23541}&\numprint{51747}&\numprint{0}&\numprint{0}&-&\checkmark&\checkmark&\checkmark&\checkmark&  \numprint{12082}&\\ 
019 &\numprint{200}&\numprint{884}&\numprint{194}&\numprint{862}&\checkmark&\checkmark&\checkmark&\checkmark&\checkmark&  \numprint{130}&\\ 
021 &\numprint{24765}&\numprint{30242}&\numprint{0}&\numprint{0}&-&\checkmark&\checkmark&\checkmark&\checkmark&  \numprint{5110}&\\ 
023 &\numprint{27717}&\numprint{133665}&\numprint{0}&\numprint{0}&-&\checkmark&\checkmark&\checkmark&\checkmark&  \numprint{16013}&\\ 
025 &\numprint{23194}&\numprint{28221}&\numprint{0}&\numprint{0}&-&\checkmark&\checkmark&\checkmark&\checkmark&  \numprint{4899}&\\ 
027 &\numprint{65866}&\numprint{81245}&\numprint{0}&\numprint{0}&-&\checkmark&\checkmark&\checkmark&\checkmark&  \numprint{13431}&\\ 
029 &\numprint{13431}&\numprint{21999}&\numprint{0}&\numprint{0}&-&\checkmark&\checkmark&\checkmark&\checkmark&  \numprint{6622}&\\ 
031 &\numprint{200}&\numprint{813}&\numprint{198}&\numprint{818}&\checkmark&\checkmark&\checkmark&\checkmark&\checkmark&  \numprint{136}&\\ 
033 &\numprint{4410}&\numprint{6885}&\numprint{138}&\numprint{471}&-&\checkmark&\checkmark&\checkmark&\checkmark&  \numprint{2725}&\\ 
035 &\numprint{200}&\numprint{884}&\numprint{189}&\numprint{859}&\checkmark&\checkmark&\checkmark&\checkmark&\checkmark&  \numprint{133}&\\ 
037 &\numprint{198}&\numprint{824}&\numprint{194}&\numprint{810}&\checkmark&\checkmark&\checkmark&\checkmark&\checkmark&  \numprint{131}&\\ 
039 &\numprint{6795}&\numprint{10620}&\numprint{219}&\numprint{753}&-&\checkmark&\checkmark&\checkmark&\checkmark&  \numprint{4200}&\\ 
041 &\numprint{200}&\numprint{1040}&\numprint{200}&\numprint{1023}&\checkmark&\checkmark&\checkmark&\checkmark&\checkmark&  \numprint{139}&\\ 
043 &\numprint{200}&\numprint{841}&\numprint{198}&\numprint{844}&\checkmark&\checkmark&\checkmark&\checkmark&\checkmark&  \numprint{139}&\\ 
045 &\numprint{200}&\numprint{1044}&\numprint{200}&\numprint{1020}&\checkmark&\checkmark&\checkmark&\checkmark&\checkmark&  \numprint{137}&\\ 
047 &\numprint{200}&\numprint{1120}&\numprint{198}&\numprint{1080}&\checkmark&\checkmark&\checkmark&\checkmark&\checkmark&  \numprint{140}&\\ 
049 &\numprint{200}&\numprint{957}&\numprint{198}&\numprint{930}&\checkmark&\checkmark&\checkmark&\checkmark&\checkmark&  \numprint{136}&\\ 
051 &\numprint{200}&\numprint{1135}&\numprint{200}&\numprint{1098}&\checkmark&\checkmark&\checkmark&\checkmark&\checkmark&  \numprint{140}&\\ 
053 &\numprint{200}&\numprint{1062}&\numprint{200}&\numprint{1026}&\checkmark&\checkmark&\checkmark&\checkmark&\checkmark&  \numprint{139}&\\ 
055 &\numprint{200}&\numprint{958}&\numprint{194}&\numprint{938}&\checkmark&\checkmark&\checkmark&\checkmark&\checkmark&  \numprint{134}&\\ 
057 &\numprint{200}&\numprint{1200}&\numprint{197}&\numprint{1139}&\checkmark&\checkmark&\checkmark&\checkmark&\checkmark&  \numprint{142}&\\ 
059 &\numprint{200}&\numprint{988}&\numprint{193}&\numprint{954}&\checkmark&\checkmark&\checkmark&\checkmark&\checkmark&  \numprint{137}&\\ 
061 &\numprint{200}&\numprint{952}&\numprint{198}&\numprint{914}&\checkmark&\checkmark&\checkmark&\checkmark&\checkmark&  \numprint{135}&\\ 
063 &\numprint{200}&\numprint{1040}&\numprint{200}&\numprint{1011}&\checkmark&\checkmark&\checkmark&\checkmark&\checkmark&  \numprint{138}&\\ 
065 &\numprint{200}&\numprint{1037}&\numprint{200}&\numprint{1011}&\checkmark&\checkmark&\checkmark&\checkmark&\checkmark&  \numprint{138}&\\ 
067 &\numprint{200}&\numprint{1201}&\numprint{200}&\numprint{1174}&\checkmark&\checkmark&\checkmark&\checkmark&\checkmark&  \numprint{143}&\\ 
069 &\numprint{200}&\numprint{1120}&\numprint{196}&\numprint{1077}&\checkmark&\checkmark&\checkmark&\checkmark&\checkmark&  \numprint{140}&\\ 
071 &\numprint{200}&\numprint{984}&\numprint{200}&\numprint{952}&\checkmark&\checkmark&\checkmark&\checkmark&\checkmark&  \numprint{136}&\\ 
073 &\numprint{200}&\numprint{1107}&\numprint{200}&\numprint{1078}&\checkmark&\checkmark&\checkmark&\checkmark&\checkmark&  \numprint{139}&\\ 
075 &\numprint{26300}&\numprint{41500}&\numprint{500}&\numprint{3000}&-&-&\checkmark&-&\checkmark&  \numprint{16300}&\\ 
077 &\numprint{200}&\numprint{988}&\numprint{193}&\numprint{954}&\checkmark&\checkmark&\checkmark&\checkmark&\checkmark&  \numprint{137}&\\ 
079 &\numprint{26300}&\numprint{41500}&\numprint{500}&\numprint{3000}&-&-&\checkmark&-&\checkmark&  \numprint{16300}&\\ 
081 &\numprint{199}&\numprint{1124}&\numprint{197}&\numprint{1087}&\checkmark&\checkmark&\checkmark&\checkmark&\checkmark&  \numprint{141}&\\ 
083 &\numprint{200}&\numprint{1215}&\numprint{198}&\numprint{1182}&\checkmark&\checkmark&\checkmark&\checkmark&\checkmark&  \numprint{144}&\\ 
085 &\numprint{11470}&\numprint{17408}&\numprint{3539}&\numprint{25955}&-&-&-&-&-&  &\\ 
087 &\numprint{13590}&\numprint{21240}&\numprint{441}&\numprint{1512}&-&\checkmark&-&-&\checkmark&  \numprint{8400}&\\ 
089 &\numprint{57316}&\numprint{77978}&\numprint{16834}&\numprint{54847}&-&-&-&-&-&  &\\ 
091 &\numprint{200}&\numprint{1196}&\numprint{200}&\numprint{1163}&\checkmark&\checkmark&\checkmark&\checkmark&\checkmark&  \numprint{145}&\\ 
093 &\numprint{200}&\numprint{1207}&\numprint{200}&\numprint{1162}&\checkmark&\checkmark&\checkmark&\checkmark&\checkmark&  \numprint{143}&\\ 
095 &\numprint{15783}&\numprint{24663}&\numprint{510}&\numprint{1746}&-&\checkmark&-&-&\checkmark&  \numprint{9755}&\\ 
097 &\numprint{18096}&\numprint{28281}&\numprint{579}&\numprint{1995}&-&\checkmark&-&-&\checkmark&  \numprint{11185}&\\ 
099 &\numprint{26300}&\numprint{41500}&\numprint{500}&\numprint{3000}&-&-&\checkmark&-&\checkmark&  \numprint{16300}&\\ 
\bottomrule
\end{tabular}
\end{table*}

\begin{table*}
\centering

\caption{Detailed per instance results for public instances. The columns $n$ and $m$ refer to the number of nodes and edges of the input graph, $n'$ and $m'$ refer to the number of nodes and edges of the kernel graph after reductions have been applied exhaustively, and $|VC|$ refers to the size of the minimum vertex cover of the input graph. We list a `\checkmark' when a solver successfully solved the given instance in the time limit, and `-' otherwise.}
\label{tab:detailedresults2}
\begin{tabular}{l@{\hskip 25pt} rrrr|ccccc|rc}
\toprule
inst\# & $n$ &$m$& $n'$& $m'$ & \AlgName{MoMC} & \AlgName{RMoMC} & \AlgName{LSBnR} & \AlgName{BnR} & \AlgName{FullA} & $|VC|$ \\
                \midrule

101 &\numprint{26300}&\numprint{41500}&\numprint{500}&\numprint{3000}&-&-&\checkmark&-&\checkmark&  \numprint{16300}&\\ 
103 &\numprint{15783}&\numprint{24663}&\numprint{513}&\numprint{1752}&-&\checkmark&-&-&\checkmark&  \numprint{9755}&\\ 
105 &\numprint{26300}&\numprint{41500}&\numprint{500}&\numprint{3000}&-&-&\checkmark&-&\checkmark&  \numprint{16300}&\\ 
107 &\numprint{13590}&\numprint{21240}&\numprint{435}&\numprint{1500}&-&\checkmark&-&-&\checkmark&  \numprint{8400}&\\ 
109 &\numprint{66992}&\numprint{90970}&\numprint{20336}&\numprint{66350}&-&-&-&-&-&  &\\ 
111 &\numprint{450}&\numprint{17831}&\numprint{450}&\numprint{17831}&\checkmark&\checkmark&-&-&\checkmark&  \numprint{420}&\\ 
113 &\numprint{26300}&\numprint{41500}&\numprint{500}&\numprint{3000}&-&-&\checkmark&-&\checkmark&  \numprint{16300}&\\ 
115 &\numprint{18096}&\numprint{28281}&\numprint{573}&\numprint{1986}&-&\checkmark&-&-&\checkmark&  \numprint{11185}&\\ 
117 &\numprint{18096}&\numprint{28281}&\numprint{582}&\numprint{2007}&-&\checkmark&-&-&\checkmark&  \numprint{11185}&\\ 
119 &\numprint{18096}&\numprint{28281}&\numprint{588}&\numprint{2016}&-&\checkmark&-&-&\checkmark&  \numprint{11185}&\\ 
121 &\numprint{18096}&\numprint{28281}&\numprint{579}&\numprint{1998}&-&\checkmark&-&-&\checkmark&  \numprint{11185}&\\ 
123 &\numprint{26300}&\numprint{41500}&\numprint{500}&\numprint{3000}&-&-&\checkmark&-&\checkmark&  \numprint{16300}&\\ 
125 &\numprint{26300}&\numprint{41500}&\numprint{500}&\numprint{3000}&-&-&\checkmark&-&\checkmark&  \numprint{16300}&\\ 
127 &\numprint{18096}&\numprint{28281}&\numprint{582}&\numprint{2001}&-&\checkmark&-&-&\checkmark&  \numprint{11185}&\\ 
129 &\numprint{15783}&\numprint{24663}&\numprint{507}&\numprint{1752}&-&\checkmark&-&-&\checkmark&  \numprint{9755}&\\ 
131 &\numprint{2980}&\numprint{5360}&\numprint{2179}&\numprint{6951}&\checkmark&-&-&-&\checkmark&  \numprint{1920}&\\ 
133 &\numprint{15783}&\numprint{24663}&\numprint{507}&\numprint{1746}&-&\checkmark&-&-&\checkmark&  \numprint{9755}&\\ 
135 &\numprint{26300}&\numprint{41500}&\numprint{500}&\numprint{3000}&-&-&\checkmark&-&\checkmark&  \numprint{16300}&\\ 
137 &\numprint{26300}&\numprint{41500}&\numprint{500}&\numprint{3000}&-&-&\checkmark&-&\checkmark&  \numprint{16300}&\\ 
139 &\numprint{18096}&\numprint{28281}&\numprint{579}&\numprint{1995}&-&\checkmark&-&-&\checkmark&  \numprint{11185}&\\ 
141 &\numprint{18096}&\numprint{28281}&\numprint{576}&\numprint{1995}&-&\checkmark&-&-&\checkmark&  \numprint{11185}&\\ 
143 &\numprint{18096}&\numprint{28281}&\numprint{582}&\numprint{2001}&-&\checkmark&-&-&\checkmark&  \numprint{11185}&\\ 
145 &\numprint{18096}&\numprint{28281}&\numprint{576}&\numprint{1989}&-&\checkmark&-&-&\checkmark&  \numprint{11185}&\\ 
147 &\numprint{18096}&\numprint{28281}&\numprint{567}&\numprint{1974}&-&\checkmark&-&-&\checkmark&  \numprint{11185}&\\ 
149 &\numprint{26300}&\numprint{41500}&\numprint{500}&\numprint{3000}&-&-&\checkmark&-&\checkmark&  \numprint{16300}&\\ 
151 &\numprint{15783}&\numprint{24663}&\numprint{501}&\numprint{1728}&-&\checkmark&-&-&\checkmark&  \numprint{9755}&\\ 
153 &\numprint{29076}&\numprint{45570}&\numprint{2124}&\numprint{16266}&-&-&-&-&-&  &\\ 
155 &\numprint{26300}&\numprint{41500}&\numprint{500}&\numprint{3000}&-&-&\checkmark&-&\checkmark&  \numprint{16300}&\\ 
157 &\numprint{2980}&\numprint{5360}&\numprint{2169}&\numprint{6898}&\checkmark&-&-&-&\checkmark&  \numprint{1920}&\\ 
159 &\numprint{18096}&\numprint{28281}&\numprint{582}&\numprint{2004}&-&\checkmark&-&-&\checkmark&  \numprint{11185}&\\ 
161 &\numprint{138141}&\numprint{227241}&\numprint{41926}&\numprint{202869}&-&-&-&-&-&  &\\ 
163 &\numprint{18096}&\numprint{28281}&\numprint{582}&\numprint{2004}&-&\checkmark&-&-&\checkmark&  \numprint{11185}&\\ 
165 &\numprint{18096}&\numprint{28281}&\numprint{576}&\numprint{1995}&-&\checkmark&-&-&\checkmark&  \numprint{11185}&\\ 
167 &\numprint{15783}&\numprint{24663}&\numprint{510}&\numprint{1746}&-&\checkmark&-&-&\checkmark&  \numprint{9755}&\\ 
169 &\numprint{4768}&\numprint{8576}&\numprint{3458}&\numprint{11014}&-&-&-&-&-&  &\\ 
171 &\numprint{18096}&\numprint{28281}&\numprint{576}&\numprint{1989}&-&\checkmark&-&-&\checkmark&  \numprint{11185}&\\ 
173 &\numprint{56860}&\numprint{77264}&\numprint{17090}&\numprint{55568}&-&-&-&-&-&  &\\ 
175 &\numprint{3523}&\numprint{6446}&\numprint{2723}&\numprint{8570}&-&-&-&-&-&  &\\ 
177 &\numprint{5066}&\numprint{9112}&\numprint{3704}&\numprint{11797}&-&-&-&-&-&  &\\ 
179 &\numprint{15783}&\numprint{24663}&\numprint{504}&\numprint{1740}&-&\checkmark&-&-&\checkmark&  \numprint{9755}&\\ 
181 &\numprint{18096}&\numprint{28281}&\numprint{573}&\numprint{1989}&-&\checkmark&\checkmark&-&\checkmark&  \numprint{11185}&\\ 
183 &\numprint{72420}&\numprint{118362}&\numprint{30340}&\numprint{133872}&-&-&-&-&-&  &\\ 
185 &\numprint{3523}&\numprint{6446}&\numprint{2723}&\numprint{8568}&-&-&-&-&-&  &\\ 
187 &\numprint{4227}&\numprint{7734}&\numprint{3264}&\numprint{10286}&-&-&-&-&-&  &\\ 
189 &\numprint{7400}&\numprint{13600}&\numprint{5802}&\numprint{18212}&-&-&-&-&-&  &\\ 
191 &\numprint{4579}&\numprint{8378}&\numprint{3539}&\numprint{11137}&-&-&-&-&-&  &\\ 
193 &\numprint{7030}&\numprint{12920}&\numprint{5510}&\numprint{17294}&-&-&-&-&-&  &\\ 
195 &\numprint{1150}&\numprint{81068}&\numprint{1150}&\numprint{81068}&-&-&-&-&-&  &\\ 
197 &\numprint{1534}&\numprint{127011}&\numprint{1534}&\numprint{127011}&-&-&-&-&-&  &\\ 
199 &\numprint{1534}&\numprint{126163}&\numprint{1534}&\numprint{126163}&-&-&-&-&-&  &\\ 

\bottomrule
\end{tabular}
\end{table*}

\begin{table*}
\centering

\caption{Detailed per instance results for private instances. The columns $n$ and $m$ refer to the number of nodes and edges of the input graph, $n'$ and $m'$ refer to the number of nodes and edges of the kernel graph after reductions have been applied exhaustively, and $|VC|$ refers to the size of the minimum vertex cover of the input graph. We list a `\checkmark' when a solver successfully solved the given instance in the time limit, and `-' otherwise.}
\label{tab:detailedresultsprivate}
\begin{tabular}{l@{\hskip 25pt} rrrr|ccccc|rc}
\toprule
inst\# & $n$ &$m$& $n'$& $m'$ & \AlgName{MoMC} & \AlgName{RMoMC} & \AlgName{LSBnR} & \AlgName{BnR} & \AlgName{FullA} & $|VC|$ \\
                \midrule
002 &\numprint{51795}&\numprint{63334}&\numprint{0}&\numprint{0}&-&\checkmark&\checkmark&\checkmark&\checkmark&  \numprint{10605}&\\ 
004 &\numprint{8114}&\numprint{26013}&\numprint{0}&\numprint{0}&-&\checkmark&\checkmark&\checkmark&\checkmark&  \numprint{2574}&\\ 
006 &\numprint{200}&\numprint{751}&\numprint{188}&\numprint{716}&\checkmark&\checkmark&\checkmark&\checkmark&\checkmark&  \numprint{126}&\\ 
008 &\numprint{7537}&\numprint{72833}&\numprint{0}&\numprint{0}&-&\checkmark&\checkmark&\checkmark&\checkmark&  \numprint{3345}&\\ 
010 &\numprint{199}&\numprint{774}&\numprint{189}&\numprint{756}&\checkmark&\checkmark&\checkmark&\checkmark&\checkmark&  \numprint{127}&\\ 
012 &\numprint{53444}&\numprint{68044}&\numprint{0}&\numprint{0}&-&\checkmark&\checkmark&\checkmark&\checkmark&  \numprint{10918}&\\ 
014 &\numprint{25123}&\numprint{31552}&\numprint{0}&\numprint{0}&-&\checkmark&\checkmark&\checkmark&\checkmark&  \numprint{5111}&\\ 
016 &\numprint{153}&\numprint{802}&\numprint{153}&\numprint{802}&-&-&-&-&-&  &\\ 
018 &\numprint{49212}&\numprint{63601}&\numprint{0}&\numprint{0}&-&\checkmark&\checkmark&\checkmark&\checkmark&  \numprint{10201}&\\ 
020 &\numprint{57287}&\numprint{71155}&\numprint{0}&\numprint{0}&-&\checkmark&\checkmark&\checkmark&\checkmark&  \numprint{11648}&\\ 
022 &\numprint{12589}&\numprint{33129}&\numprint{0}&\numprint{0}&-&\checkmark&\checkmark&\checkmark&\checkmark&  \numprint{6749}&\\ 
024 &\numprint{7620}&\numprint{47293}&\numprint{0}&\numprint{0}&-&\checkmark&\checkmark&\checkmark&\checkmark&  \numprint{4364}&\\ 
026 &\numprint{6140}&\numprint{36767}&\numprint{0}&\numprint{0}&-&\checkmark&\checkmark&\checkmark&\checkmark&  \numprint{2506}&\\ 
028 &\numprint{54991}&\numprint{67000}&\numprint{0}&\numprint{0}&-&\checkmark&\checkmark&\checkmark&\checkmark&  \numprint{11211}&\\ 
030 &\numprint{62853}&\numprint{79557}&\numprint{0}&\numprint{0}&-&\checkmark&\checkmark&\checkmark&\checkmark&  \numprint{13338}&\\ 
032 &\numprint{1490}&\numprint{2680}&\numprint{1081}&\numprint{3426}&\checkmark&-&-&-&\checkmark&  \numprint{960}&\\ 
034 &\numprint{1490}&\numprint{2680}&\numprint{1090}&\numprint{3467}&\checkmark&\checkmark&-&-&\checkmark&  \numprint{960}&\\ 
036 &\numprint{26300}&\numprint{41500}&\numprint{500}&\numprint{3000}&-&\checkmark&\checkmark&\checkmark&\checkmark&  \numprint{16300}&\\ 
038 &\numprint{786}&\numprint{14024}&\numprint{460}&\numprint{6623}&\checkmark&\checkmark&\checkmark&\checkmark&\checkmark&  \numprint{605}&\\ 
040 &\numprint{210}&\numprint{625}&\numprint{210}&\numprint{625}&\checkmark&\checkmark&-&-&\checkmark&  \numprint{145}&\\ 
042 &\numprint{200}&\numprint{974}&\numprint{200}&\numprint{952}&\checkmark&\checkmark&\checkmark&\checkmark&\checkmark&  \numprint{136}&\\ 
044 &\numprint{200}&\numprint{1186}&\numprint{200}&\numprint{1147}&\checkmark&\checkmark&\checkmark&\checkmark&\checkmark&  \numprint{142}&\\ 
046 &\numprint{200}&\numprint{812}&\numprint{200}&\numprint{812}&\checkmark&\checkmark&\checkmark&\checkmark&\checkmark&  \numprint{137}&\\ 
048 &\numprint{200}&\numprint{1052}&\numprint{198}&\numprint{1022}&\checkmark&\checkmark&\checkmark&\checkmark&\checkmark&  \numprint{138}&\\ 
050 &\numprint{200}&\numprint{1048}&\numprint{200}&\numprint{1025}&\checkmark&\checkmark&\checkmark&\checkmark&\checkmark&  \numprint{140}&\\ 
052 &\numprint{200}&\numprint{1019}&\numprint{198}&\numprint{1000}&\checkmark&\checkmark&\checkmark&\checkmark&\checkmark&  \numprint{138}&\\ 
054 &\numprint{200}&\numprint{985}&\numprint{198}&\numprint{951}&\checkmark&\checkmark&\checkmark&\checkmark&\checkmark&  \numprint{137}&\\ 
056 &\numprint{200}&\numprint{1117}&\numprint{200}&\numprint{1089}&\checkmark&\checkmark&\checkmark&\checkmark&\checkmark&  \numprint{141}&\\ 
058 &\numprint{200}&\numprint{1202}&\numprint{200}&\numprint{1171}&\checkmark&\checkmark&\checkmark&\checkmark&\checkmark&  \numprint{142}&\\ 
060 &\numprint{200}&\numprint{1147}&\numprint{200}&\numprint{1118}&\checkmark&\checkmark&\checkmark&\checkmark&\checkmark&  \numprint{141}&\\ 
062 &\numprint{199}&\numprint{1164}&\numprint{199}&\numprint{1128}&\checkmark&\checkmark&\checkmark&\checkmark&\checkmark&  \numprint{141}&\\ 
064 &\numprint{200}&\numprint{1071}&\numprint{198}&\numprint{1040}&\checkmark&\checkmark&\checkmark&\checkmark&\checkmark&  \numprint{138}&\\ 
066 &\numprint{200}&\numprint{884}&\numprint{198}&\numprint{875}&\checkmark&\checkmark&\checkmark&\checkmark&\checkmark&  \numprint{134}&\\ 
068 &\numprint{200}&\numprint{983}&\numprint{198}&\numprint{961}&\checkmark&\checkmark&\checkmark&\checkmark&\checkmark&  \numprint{135}&\\ 
070 &\numprint{200}&\numprint{887}&\numprint{198}&\numprint{856}&\checkmark&\checkmark&\checkmark&\checkmark&\checkmark&  \numprint{133}&\\ 
072 &\numprint{200}&\numprint{1204}&\numprint{198}&\numprint{1176}&\checkmark&\checkmark&\checkmark&\checkmark&\checkmark&  \numprint{140}&\\ 
074 &\numprint{200}&\numprint{820}&\numprint{194}&\numprint{785}&\checkmark&\checkmark&\checkmark&\checkmark&\checkmark&  \numprint{132}&\\ 
076 &\numprint{26300}&\numprint{41500}&\numprint{500}&\numprint{3000}&-&\checkmark&\checkmark&-&\checkmark&  \numprint{16300}&\\ 
078 &\numprint{11349}&\numprint{17739}&\numprint{357}&\numprint{1245}&-&\checkmark&-&-&\checkmark&  \numprint{7015}&\\ 
080 &\numprint{26300}&\numprint{41500}&\numprint{500}&\numprint{3000}&-&-&\checkmark&-&\checkmark&  \numprint{16300}&\\ 
082 &\numprint{200}&\numprint{978}&\numprint{196}&\numprint{956}&\checkmark&\checkmark&\checkmark&\checkmark&\checkmark&  \numprint{138}&\\ 
084 &\numprint{13590}&\numprint{21240}&\numprint{435}&\numprint{1503}&-&\checkmark&-&-&\checkmark&  \numprint{8400}&\\ 
086 &\numprint{26300}&\numprint{41500}&\numprint{500}&\numprint{3000}&-&\checkmark&\checkmark&-&\checkmark&  \numprint{16300}&\\ 
088 &\numprint{26300}&\numprint{41500}&\numprint{500}&\numprint{3000}&-&\checkmark&\checkmark&-&\checkmark&  \numprint{16300}&\\ 
090 &\numprint{11349}&\numprint{17739}&\numprint{357}&\numprint{1245}&-&\checkmark&-&-&\checkmark&  \numprint{7015}&\\ 
092 &\numprint{450}&\numprint{17794}&\numprint{450}&\numprint{17794}&\checkmark&\checkmark&-&-&\checkmark&  \numprint{420}&\\ 
094 &\numprint{5960}&\numprint{10720}&\numprint{4217}&\numprint{13456}&-&-&-&-&-&  &\\ 
096 &\numprint{26300}&\numprint{41500}&\numprint{500}&\numprint{3000}&-&-&\checkmark&-&\checkmark&  \numprint{16300}&\\ 
098 &\numprint{26300}&\numprint{41500}&\numprint{500}&\numprint{3000}&-&-&\checkmark&-&\checkmark&  \numprint{16300}&\\ 
100 &\numprint{26300}&\numprint{41500}&\numprint{500}&\numprint{3000}&-&\checkmark&\checkmark&\checkmark&\checkmark&  \numprint{16300}&\\ 
\bottomrule
\end{tabular}
\end{table*}

\begin{table*}
\centering

\caption{Detailed per instance results for private instances. The columns $n$ and $m$ refer to the number of nodes and edges of the input graph, $n'$ and $m'$ refer to the number of nodes and edges of the kernel graph after reductions have been applied exhaustively, and $|VC|$ refers to the size of the minimum vertex cover of the input graph. We list a `\checkmark' when a solver successfully solved the given instance in the time limit, and `-' otherwise.}
\label{tab:detailedresults2private}
\begin{tabular}{l@{\hskip 25pt} rrrr|ccccc|rc}
\toprule
inst\# & $n$ &$m$& $n'$& $m'$ & \AlgName{MoMC} & \AlgName{RMoMC} & \AlgName{LSBnR} & \AlgName{BnR} & \AlgName{FullA} & $|VC|$ \\
                \midrule
102 &\numprint{26300}&\numprint{41500}&\numprint{500}&\numprint{3000}&-&-&\checkmark&-&\checkmark&  \numprint{16300}&\\ 
104 &\numprint{26300}&\numprint{41500}&\numprint{500}&\numprint{3000}&-&-&\checkmark&-&\checkmark&  \numprint{16300}&\\ 
106 &\numprint{2980}&\numprint{5360}&\numprint{2136}&\numprint{6809}&\checkmark&-&-&-&\checkmark&  \numprint{1920}&\\ 
108 &\numprint{26300}&\numprint{41500}&\numprint{500}&\numprint{3000}&-&-&\checkmark&-&\checkmark&  \numprint{16300}&\\ 
110 &\numprint{98128}&\numprint{161357}&\numprint{29168}&\numprint{140392}&-&-&-&-&-&  &\\ 
112 &\numprint{18096}&\numprint{28281}&\numprint{576}&\numprint{1992}&-&\checkmark&-&-&\checkmark&  \numprint{11185}&\\ 
114 &\numprint{15783}&\numprint{24663}&\numprint{504}&\numprint{1740}&-&\checkmark&-&-&\checkmark&  \numprint{9755}&\\ 
116 &\numprint{26300}&\numprint{41500}&\numprint{500}&\numprint{3000}&-&-&\checkmark&-&\checkmark&  \numprint{16300}&\\ 
118 &\numprint{26300}&\numprint{41500}&\numprint{500}&\numprint{3000}&-&-&\checkmark&-&\checkmark&  \numprint{16300}&\\ 
120 &\numprint{70144}&\numprint{116378}&\numprint{6029}&\numprint{38285}&-&-&-&-&-&  &\\ 
122 &\numprint{26300}&\numprint{41500}&\numprint{500}&\numprint{3000}&-&-&\checkmark&-&\checkmark&  \numprint{16300}&\\ 
124 &\numprint{26300}&\numprint{41500}&\numprint{500}&\numprint{3000}&-&-&\checkmark&-&\checkmark&  \numprint{16300}&\\ 
126 &\numprint{18096}&\numprint{28281}&\numprint{582}&\numprint{2001}&-&\checkmark&-&-&\checkmark&  \numprint{11185}&\\ 
128 &\numprint{26300}&\numprint{41500}&\numprint{500}&\numprint{3000}&-&-&-&-&-&  &\\ 
130 &\numprint{26300}&\numprint{41500}&\numprint{500}&\numprint{3000}&-&-&\checkmark&-&\checkmark&  \numprint{16300}&\\ 
132 &\numprint{15783}&\numprint{24663}&\numprint{513}&\numprint{1755}&-&\checkmark&-&-&\checkmark&  \numprint{9755}&\\ 
134 &\numprint{26300}&\numprint{41500}&\numprint{500}&\numprint{3000}&-&-&\checkmark&-&\checkmark&  \numprint{16300}&\\ 
136 &\numprint{18096}&\numprint{28281}&\numprint{585}&\numprint{2007}&-&\checkmark&-&-&\checkmark&  \numprint{11185}&\\ 
138 &\numprint{18096}&\numprint{28281}&\numprint{576}&\numprint{1992}&-&\checkmark&-&-&\checkmark&  \numprint{11185}&\\ 
140 &\numprint{26300}&\numprint{41500}&\numprint{500}&\numprint{3000}&-&-&\checkmark&-&\checkmark&  \numprint{16300}&\\ 
142 &\numprint{2980}&\numprint{5360}&\numprint{2180}&\numprint{6946}&\checkmark&-&-&-&\checkmark&  \numprint{1920}&\\ 
144 &\numprint{26300}&\numprint{41500}&\numprint{500}&\numprint{3000}&-&-&\checkmark&-&\checkmark&  \numprint{16300}&\\ 
146 &\numprint{26300}&\numprint{41500}&\numprint{500}&\numprint{3000}&-&-&\checkmark&-&\checkmark&  \numprint{16300}&\\ 
148 &\numprint{26300}&\numprint{41500}&\numprint{500}&\numprint{3000}&-&-&\checkmark&-&\checkmark&  \numprint{16300}&\\ 
150 &\numprint{26300}&\numprint{41500}&\numprint{500}&\numprint{3000}&-&-&\checkmark&-&\checkmark&  \numprint{16300}&\\ 
152 &\numprint{13590}&\numprint{21240}&\numprint{438}&\numprint{1506}&-&\checkmark&\checkmark&-&\checkmark&  \numprint{8400}&\\ 
154 &\numprint{15783}&\numprint{24663}&\numprint{504}&\numprint{1737}&-&\checkmark&-&-&\checkmark&  \numprint{9755}&\\ 
156 &\numprint{450}&\numprint{17809}&\numprint{450}&\numprint{17809}&\checkmark&\checkmark&-&-&\checkmark&  \numprint{420}&\\ 
158 &\numprint{15783}&\numprint{24663}&\numprint{507}&\numprint{1746}&-&\checkmark&-&-&\checkmark&  \numprint{9755}&\\ 
160 &\numprint{18096}&\numprint{28281}&\numprint{576}&\numprint{1989}&-&\checkmark&-&-&\checkmark&  \numprint{11185}&\\ 
162 &\numprint{50635}&\numprint{83075}&\numprint{13066}&\numprint{63758}&-&-&-&-&-&  &\\ 
164 &\numprint{29296}&\numprint{46040}&\numprint{1210}&\numprint{8666}&-&-&-&-&-&  &\\ 
166 &\numprint{3278}&\numprint{5896}&\numprint{2400}&\numprint{7643}&\checkmark&-&-&-&-& \numprint{2112} &\\ 
168 &\numprint{2980}&\numprint{5360}&\numprint{2180}&\numprint{6943}&\checkmark&-&-&-&\checkmark&  \numprint{1920}&\\ 
170 &\numprint{15783}&\numprint{24663}&\numprint{507}&\numprint{1746}&-&\checkmark&-&-&\checkmark&  \numprint{9755}&\\ 
172 &\numprint{4025}&\numprint{7435}&\numprint{3158}&\numprint{9863}&-&-&-&-&-&  &\\ 
174 &\numprint{2980}&\numprint{5360}&\numprint{2180}&\numprint{6955}&\checkmark&-&-&-&\checkmark&  \numprint{1920}&\\ 
176 &\numprint{15783}&\numprint{24663}&\numprint{501}&\numprint{1734}&-&\checkmark&-&-&\checkmark&  \numprint{9755}&\\ 
178 &\numprint{18096}&\numprint{28281}&\numprint{573}&\numprint{1995}&-&\checkmark&-&-&\checkmark&  \numprint{11185}&\\ 
180 &\numprint{15783}&\numprint{24663}&\numprint{501}&\numprint{1731}&-&\checkmark&-&-&\checkmark&  \numprint{9755}&\\ 
182 &\numprint{26300}&\numprint{41500}&\numprint{500}&\numprint{3000}&-&-&\checkmark&-&\checkmark&  \numprint{16300}&\\ 
184 &\numprint{6290}&\numprint{11560}&\numprint{4904}&\numprint{15397}&-&-&-&-&-&  &\\ 
186 &\numprint{26300}&\numprint{41500}&\numprint{500}&\numprint{3000}&-&-&\checkmark&-&\checkmark&  \numprint{16300}&\\ 
188 &\numprint{6660}&\numprint{12240}&\numprint{5220}&\numprint{16375}&-&-&-&-&-&  &\\ 
190 &\numprint{3875}&\numprint{7090}&\numprint{2997}&\numprint{9424}&-&-&-&-&-&  &\\ 
192 &\numprint{2980}&\numprint{5360}&\numprint{2180}&\numprint{6941}&\checkmark&-&-&-&\checkmark&  \numprint{1920}&\\ 
194 &\numprint{1150}&\numprint{80851}&\numprint{1150}&\numprint{80851}&\checkmark&\checkmark&-&-&\checkmark&  \numprint{1100}&\\ 
196 &\numprint{1534}&\numprint{126082}&\numprint{1534}&\numprint{126082}&-&-&-&-&-&  &\\ 
198 &\numprint{1150}&\numprint{80072}&\numprint{1150}&\numprint{80072}&\checkmark&\checkmark&-&-&\checkmark&  \numprint{1100}&\\ 
200 &\numprint{1150}&\numprint{80258}&\numprint{1150}&\numprint{80258}&\checkmark&\checkmark&-&-&\checkmark&  \numprint{1100}&\\
\bottomrule
\end{tabular}
\end{table*}
\end{document}